\documentclass[%
 reprint,
 amsmath,amssymb,
 aps,
longbibliography
]{revtex4-2}

\usepackage{graphicx}
\usepackage{dcolumn}
\usepackage{bm}
\usepackage{mathrsfs} 
\usepackage{tikz}
\usepackage[colorlinks=true,allcolors=black]{hyperref}

\begin{document}

\preprint{APS/123-QED}

\title{A generating-function approach to the interference \\ of squeezed states with partial distinguishability}

\author{Matheus Eiji Ohno Bezerra}
\author{Valery Shchesnovich}
\affiliation{Center for Natural and Human Sciences, Universidade Federal do ABC, 09210-170 Santo Andr\'e, Brazil}

\date{\today}

\begin{abstract}

Photon distinguishability is a fundamental property manifested in multiphoton interference and one of the main sources of noise in any photonic quantum information processing.  In this work, rather than relying on first-quantization methods, we build on a generating-function framework based on the phase-space formalism to characterize the effects of partial distinguishability on the interference of single-mode squeezed states. Our approach goes beyond commonly used models that represent distinguishability via additional noninterfering modes and captures genuine multiphoton interference effects induced by the overlap of the internal state of the photons. This description provides a clear physical account of how distinguishability gives rise to effective noise in Gaussian boson sampling protocols while enabling a systematic investigation of phase effects arising from the overlap of the internal states.

\end{abstract}

\maketitle

\section{\label{sec:int} Introduction}



Photon distinguishability originates from imperfect overlap of the internal states and constitutes a major source of noise in any photonic quantum information processing by degrading the multiphoton interference required for quantum computational advantage~\cite{ValeryBS1, BSAlg1,ValeryPRA2019,RenemaSimulability,RenemaSimulabilityGBS,Shi2022, QuesadaLoopHafnian}. The internal state of each photon is characterized by unresolved degrees of freedom, such as its temporal wave packet, spectral profile, and polarization~\cite{HOM1987,Loudon1989,Mandel1991,OuTempDist, 4photon}. Its basic manifestation is the Hong-Ou-Mandel (HOM) effect \cite{HOM1987}, where destructive two-photon interference at a balanced beamsplitter suppresses coincidence events for indistinguishable photons, which is degraded as distinguishability is introduced. For multiphoton Fock-state interference, partial distinguishability exhibits a richer structure, including collective phase effects \cite{CollectivePhase2,CollectivePhase1,CollectivePhase3}.


In the context of optical quantum computing, partial distinguishability has been extensively investigated in Fock state boson sampling (FBS) \cite{ValeryBS2, TichyBS} and identified as one of the main sources of noise capable of compromising the quantum advantage of such architectures \cite{ValeryBS1,BSAlg1,ValeryPRA2019,RenemaSimulability}. Among the various boson-sampling architectures, the Gaussian boson sampling (GBS) \cite{GBS} employs single-mode squeezed states as the input states. Due to the efficient generation of squeezed states via second- \cite{GenerationSMSS1,GenerationSMSS2} and third-order \cite{GenerationSMSS3} nonlinear optical processes, GBS offers favorable scalability in both the number of modes and the average photon number. This scalability has enabled recent large-scale experimental implementations, which reported evidence of quantum computational advantage ~\cite{Pan2020,Pan2023,Pan2025,Madsen2022,Paesani2019}. As in FBS, partial distinguishability is also one of the most significant sources of noise in GBS, degrading multiphoton quantum interference and therefore compromising the quantum advantage \cite{RenemaSimulabilityGBS,Shi2022,RenemaSimulability}.

However, in contrast to Fock-state interference, the effects of partial distinguishability in squeezed-state interference have not yet been fully explored. Previous works have addressed distinguishability using virtual-mode \cite{Shi2022,GBSBinnedValidation} and coarse-grained \cite{QuesadaLoopHafnian} models, which are well suited for validation and simulability investigations. At the same time, such approaches do not capture the full structure of partial distinguishability effects in squeezed-state interference. While this problem can also be formulated in a general way using a first-quantization framework~\cite{RenemaSimulabilityGBS,ValeryGBS}, this description is not the most convenient when dealing with Gaussian states. From a complementary physical perspective, partial distinguishability has also been shown to arise from the intrinsic spectral mixedness of photons generated from the same source, being quantified by the Schmidt number of the spectral correlations~\cite{Christ_2011}. In this work, we investigate the interference of squeezed states using the phase-space formalism, providing a natural and clear framework to treat partial distinguishability, quantified by the widely used overlap matrix of the internal states~\cite{dist_GramMatrix,ValeryBS2, TichyBS}.

The paper is organized as follows. In Sec.~\ref{sec:GenFunc}, we revisit key results from Ref.~\cite{Valery2017} and extend them to the interference of single-mode squeezed states, deriving a general expression for the output probabilities of squeezed-state interference with partially distinguishable photons in arbitrary linear interferometers. We then recover the indistinguishable-photon limit and, finally, derive the corresponding expressions for threshold detectors. Following this, we present two applications. In Sec.~\ref{sec:application1}, we reduce our results to the simple model where the photons have homogeneous overlap, providing a clear physical interpretation of the noise effects emerging from partial distinguishability. Finally, in Sec.~\ref{sec:application2}, we investigate partial distinguishability from a more fundamental perspective,
including zero-probability events and the phase effects emerging from the internal state overlap.


\section{\label{sec:GenFunc} Probability events and generating function}

\begin{figure}[t]
    \centering
\includegraphics[width=0.7 \columnwidth]{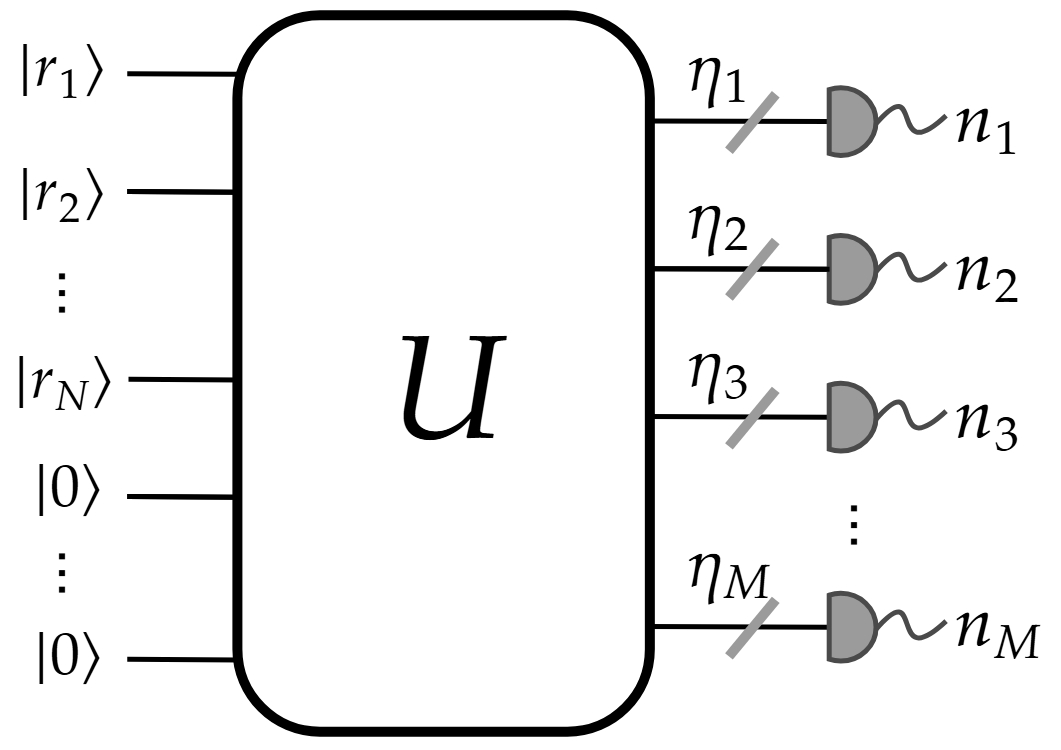}
\caption{
We consider an $M$-mode multiport
described by a unitary matrix $U$. Each input mode $k$ is prepared in a single-mode squeezed vacuum state $|r_k\rangle$, as defined in Eq.~(\ref{smss_input}). These states are injected
into the multiport, which implements a linear transformation on the spatial modes according to Eq.~(\ref{spatial_transf}). At the output, photon-number-resolving detectors with efficiencies $\eta_l$ measure the number of photons $n_l$ in each output mode.}
\label{scheme_smss}
\end{figure}

We consider an $M$-mode linear optical interferometer, or multiport, described by a unitary matrix $U$. Each input mode $k = 1, \ldots, M$ is injected with a single-mode squeezed vacuum state $|r_k\rangle$, which we refer to simply as a squeezed state, as illustrated in Fig.~\ref{scheme_smss}. We assume that the photons
generated by the source in input mode $k$ occupy an internal state
$|\psi_k\rangle$. The overall $M$-mode input state is therefore given by
$|{\bf r}\rangle = \bigotimes_{k=1}^M |r_k\rangle$, where the state in each mode is given by
\begin{equation}
    | r_k \rangle = \left(1-|S_k|^2\right)^{\frac{1}{4}} \, \text{exp} \left[ \frac{S_k}{2} ( \hat{a}^\dagger_{k, \psi_k} )^2 \right] |0 \rangle ,
    \label{smss_input}
\end{equation}
where $S_k = \tanh r_k\, e^{i\theta_k}$, with $r_k$ denoting the squeezing parameter and $\theta_k$ the squeezing phase. As commonly assumed in the GBS, we consider that the first $N$ modes of the interferometer are injected with
squeezed light, while the remaining $M-N$ modes are prepared in the vacuum state, by setting $r_k = 0$ for $k = N+1,\ldots,M$. The operator $\hat{a}^\dagger_{k,\psi_k}$ denotes the creation operator
associated with input spatial mode $k$, where the first index labels the spatial optical mode of the interferometer and the second index labels the corresponding internal state of the photon. The action of the multiport is described by a unitary transformation that mixes the spatial modes, mapping the input creation operators onto the output creation operators $\hat{b}^\dagger_{l,\psi_k}$ according to the linear transformation
\begin{equation}
    \hat{a}^\dagger_{k,\psi_k} = \sum^M_{l=1} U_{kl} \, \hat{b}^\dagger_{l,\psi_k},
    \label{spatial_transf}
\end{equation}
while leaving the internal degrees of freedom unchanged. We also introduce an orthonormal basis for the internal mode, such that the creation operators associated with the internal states can be spanned as
\begin{equation}
    \hat{a}^\dagger_{k,\psi_k} = \sum^D_{s=1} \phi_{ks} \, \hat{a}^\dagger_{k,s} ,
    \label{internal_transf}
 \end{equation}
where $D$ denotes the dimension of the internal-state Hilbert space. From Eq.~(\ref{internal_transf}), we define the overlap matrix of the internal state of the photons as the corresponding Gram matrix, which provides a standard quantitative measure of partial
distinguishability~\cite{dist_GramMatrix,ValeryBS2, TichyBS},
 \begin{equation}
     V_{kj} \equiv \langle \psi_k | \psi_j \rangle = \sum^D_{s=1} \phi^*_{ks} \phi_{js} .
 \end{equation} 

For an output configuration ${\bf n}=(n_1,\ldots,n_M)$, we employ the standard Poissonian detection operator to express the corresponding output probability \cite{BookMandel}:
\begin{equation}
    P({\bf n}) = 
    \left\langle \mathcal{N} \left\{ \prod_{l=1}^M \frac{\hat N_l^{\,n_l}}{n_l!}\,e^{-\hat N_l} \right\} \right\rangle ,
    \label{prob_Poissonian}
\end{equation}
where $\mathcal{N}$ reads for the normal ordering of creation and annihilation operators. Here we introduce the detector efficiencies as $0 \leq \eta_l \leq 1$ and define the photon-number operator as 
\begin{equation}
    \hat{N}_l = \eta_l \sum^D_{s=1} \hat{b}^\dagger_{l,s} \hat{b}_{l,s},
    \label{photon_number_def}
\end{equation}
where the summation over the internal-state basis is performed because the detectors are insensitive to internal degrees of freedom of the photons. Thus, introducing the efficiency vector ${\bm \eta} = (\eta_1, ... ,\eta_M)$, the probability can be recast in terms of a generating function as
\begin{equation}
    P({\bf n}) = \left. \prod^M_{l=1} \frac{\eta^{n_l}_l}{n_l!} \left( - \frac{\partial}{\partial \eta_l} \right)^{n_l} G({\bm \eta}) \right|_{{\bm \eta} = {\bm 1}} ,
    \label{probability_gen_func}
\end{equation}
where we assume full efficient detectors, evaluating the derivatives at ${\bm 1} = (1,\ldots,1)$, and the generating function is defined as
\begin{equation}
    G({\bm \eta}) \equiv \left\langle \mathcal{N} \left\{ \prod^M_{l=1} \text{e}^{-\hat{N}_l} \right\} \right\rangle  . 
    \label{vacuum_prob}
\end{equation}

Recent works have also employed generating functions formally equivalent to that in Eq.~(\ref{vacuum_prob}) in both FBS \cite{BSBinnedValidation,BSBinnedValidation_exp} and GBS \cite{GBSBinnedValidation,QuesadaMoments,QuesadaLoopHafnian}. In the latter case, the partial distinguishability is incorporated through a virtual-mode approach in Ref.~\cite{GBSBinnedValidation} and through a coarse-grained model in Ref.~\cite{QuesadaLoopHafnian}, which differs from our approach. 

At a more fundamental level, a central result derived by Shchesnovich in Ref.~\cite{Valery2017} provides an explicit phase-space representation of the generating function in Eq.~(\ref{vacuum_prob}), valid for arbitrary internal-state overlap. The relevant details of this formulation are also summarized in Appendix~\ref{appendix_gen_func_smss}. Introducing the phase-space coordinates ${\bm \alpha} = (\alpha_1,\ldots,\alpha_M)$ and the denoting the integral $\int d^2{\bm \alpha} \equiv \int d^2\alpha_1 \cdots \int d^2\alpha_M$, the expectation value in Eq.~(\ref{vacuum_prob}) can be evaluated using the corresponding Husimi function $Q({\bm \alpha})$, resulting in the following expression
\cite{Valery2017}
\begin{equation}
    G({\bm \eta}) = \int d^2{\bm \alpha} \, Q({\bm \alpha}) \, \frac{\text{exp}\left[- {\bm \alpha}^\dagger \left(H^{-1} - I_M \right) {\bm \alpha} \right]}{\text{det}(H)} ,
    \label{gen_func_definition}
\end{equation}
where we define the following Hermitian matrix:
\begin{equation}
    H = U \Lambda U^\dagger \circ V .
    \label{def_H}
\end{equation}
Here, $\Lambda_{ij} = (1-\eta_i)\delta_{ij}$ is a diagonal matrix encoding the detector efficiencies, and ``$\circ$'' denotes the Hadamard (element-wise) product. 

Introducing the complete set of phase-space coordinates $\mathcal{A} =(\alpha_1,\ldots,\alpha_M,\alpha_1^{*},\ldots,\alpha_M^{*})$,
the corresponding Husimi function of the squeezed states defined in Eq.~(\ref{smss_input}) reads:
\begin{eqnarray}
    Q_{\bf r}({\bm \alpha}) = \frac{c_{\bf r}}{\pi^M}~\text{exp} \left[- \frac{1}{2} \mathcal{A}^\dagger 
    \begin{pmatrix}
        I_M & \overline{D}_{\bf r} \\
        D_{\bf r} & I_M
    \end{pmatrix}
    \mathcal{A} \, \right] ,
    \label{hussimi_smss}
\end{eqnarray}
where $c_{\bf r} = \prod_{k=1}^{M} \sqrt{1-|S_k|^2}$ is a normalization coefficient and $D_{\bf r} = \mathrm{diag}(S_1,\ldots,S_M)$ is a diagonal matrix encoding the squeezing contributions. Finally, substituting Eq.~(\ref{hussimi_smss}) into Eq.~(\ref{gen_func_definition}), we obtain the following integral
\begin{align}
    G_{\bf r}({\bm \eta}) 
    &= \frac{c_{\bf r}}{\text{det}(H)} \int \frac{d^2{\bm \alpha}}{\pi^M}
    \nonumber\\
    & \quad \times \, \text{exp}\left[ - \frac{1}{2} \mathcal{A}^\dagger 
    \begin{pmatrix}
        H^{-1} & \overline{D}_{\bf r} \\
        D_{\bf r} & \overline{H}^{-1}
    \end{pmatrix}
    \mathcal{A} \right]
    ,
    \label{gen_func_smss_integral}
\end{align}
and carrying out the calculations detailed in Appendix~\ref{appendix_gen_func_smss} we arrive in the final expression of the generating function
\begin{equation}
    G_{\bf r}({\bm \eta}) 
    = c_{\bf r} \, \text{det}
    \left( I_{2M} - \mathcal{M}_{\bf r} \right)^{-\frac{1}{2}},
    \label{gen_func_smss} 
\end{equation}
where we have introduced $H_{\bf r}=D^{\frac{1}{2}}_{\bf r} H \overline{D}^{\frac{1}{2}}_{\bf r}$ and defined the Hermitian matrix:
\begin{equation}
    \mathcal{M}_{\bf r}=
    \begin{pmatrix}
    0_M & H_{\bf r}\\
    \overline{H}_{\bf r} & 0_M
    \end{pmatrix}.
    \label{H_r}
\end{equation}


\subsection{\label{subsec:hafnian} Output probability for indistinguishable photons and the relation with the hafnian}

For consistency, we reduce our general result to the limit of completely indistinguishable photons in order to establish aconnection with previous results. In this limit, the output probability is expressed in terms of a matrix hafnian, which lies at the core of the computational hardness of Gaussian boson sampling~\cite{GBS,GBSdetailed}. By making use of the Hafnian Master Theorem recently proven in Ref.~\cite{HafnianMasterTheorem}, we recover the standard hafnian expression for the output probabilities, also in agreement with the generating function derived in Ref.~\cite{QuesadaMoments}. In this way, we assume that all photons occupy the same internal state, so that the overlap matrix satisfies $V_{ij}=1$ for all $i,j$. Consequently, the Hermitian matrix in Eq.~(\ref{H_r}) becomes
\begin{equation}
    \mathcal{M}^0_{\bf r}=
    \begin{pmatrix}
    0_M & H^0_{\bf r}\\
    \overline{H}^0_{\bf r} & 0_M
    \end{pmatrix}
    ,
    \quad
    H^0_{\bf r} \equiv \big. H_{\bf r} \big|_{V_{ij}=1}
    ,
    \label{H_r_ind}
\end{equation}
which follows that $H^0_{\bf r} = D^{\frac{1}{2}}_{\bf r} U \Lambda U^\dagger \overline{D}^{\frac{1}{2}}_{\bf r}$.

After properly decomposing $H_{\bf r}^{0}$ in terms of the matrices $\Lambda$, $U$, and $D_{\bf r}$, and reordering the matrices inside the determinant using Sylvester's identity, we can rewrite the determinant in Eq.~(\ref{gen_func_smss}) as
\begin{align}
    \text{det}
    (I_{2M} - \mathcal{M}^0_{\bf r}) 
    & =
    \text{det}
    \left[ I_{2M} -
    \begin{pmatrix}
        0_M &  \Lambda \\
        \Lambda & 0_M
    \end{pmatrix}
    \begin{pmatrix}
        B & 0_M \\
        0_M & \overline{B}
    \end{pmatrix} 
    \right] 
    \label{gen_func_ind_aux}
\end{align}
where in the last line we introduced the symmetric matrix $B = U^t D_{\bf r} U$, which recovers the generating function reported in Ref.~\cite{QuesadaMoments} with a different parametrization and zero displacement. Following the Hafnian Master Theorem~\cite{HafnianMasterTheorem}, we recognize in
Eq.~(\ref{gen_func_ind_aux}) the generating-function structure corresponding to the hafnian of the matrix $B_{\mathbf n}\oplus\overline{B}_{\mathbf n}$,
\begin{equation}
    \text{det}
    (I_{2M} - \mathcal{M}^0_{\bf r})^{-\frac{1}{2}}  =  \sum_{\bf n} \big| \text{haf}(B_{\bf n}) \big|^2 \prod^M_{k=1} \frac{(1-\eta_k)}{n_k!}^{n_k} ,
    \label{gen_func_ind}
\end{equation}
where the $|{\bf n}|\times |{\bf n}|$ matrix $B_{\bf n}$ is obtained from the $M\times M$ matrix $B$ by repeating each $l$-th row and column $n_l$ times. Additionally, within the coarse-grained model of distinguishability, the internal modes can be partitioned into interfering and noninterfering subspaces, such that the resulting output probabilities are expressed in terms of a blocked loop hafnian~\cite{QuesadaLoopHafnian}.

Finally, by inverting Eq.~(\ref{gen_func_ind}) and using the general expression
for the probability in Eq.~(\ref{probability_gen_func}), we recover the standard GBS probability for indistinguishable photons with fully efficient detectors~\cite{GBS,GBSdetailed},
\begin{align}
    P^0_{\bf r} ({\bf n})
    &= c_{\bf r} \left. \prod^M_{k=1} \frac{\eta^{n_k}_k}{n_k!} \left( - \frac{\partial}{\partial \eta_k} \right)^{n_k} \text{det}
    (I_{2M} - \mathcal{M}^0_{\bf r})^{-\frac{1}{2}} \right|_{{\bm \eta} = {\bm 1}} 
    \nonumber\\
    & = \frac{c_{\bf r}}{\prod^M_{k=1}n_k!} \, \big| \text{haf}(B_{\bf n}) \big|^2 .
    \label{probability_hafnian}
\end{align}


\subsection{\label{subsec:ClickProb} Threshold detectors}

As an alternative to the photon-number-resolving detection considered above, we also consider threshold detectors, which gives a binary outcome: either no photons are detected (no-click) or one or more photons are detected (click). Such a scheme was first proposed in Ref.~\cite{QuesadaGBSClick} and has since been adopted as the detection strategy in the main experimental demonstrations of Gaussian boson sampling~\cite{Pan2020,Pan2023}. For perfectly indistinguishable photons, the corresponding output probabilities are expressed in terms of the matrix function known as the torontonian, which is believed to be classically hard to compute in the low-collision regime ~\cite{QuesadaGBSClick}. In the following, we show that, using our method, we can express the probability of such click events in terms of the generating function defined in Eq.~(\ref{gen_func_smss}).

To begin, denoting the vacuum POVM element in the output mode $k$ as
\begin{equation}
    \hat{\Pi}_{0_k}=|0_k\rangle\langle 0_k|
    ,
    \label{probability_no_click}
\end{equation}
it follows directly from
Eqs.~(\ref{probability_gen_func}) and~(\ref{vacuum_prob}) that the probability of detecting no photon in the corresponding output mode is given by
\begin{equation}
    P_{\bf r}(0_k) \equiv \langle \hat{\Pi}_{0_k}  \rangle = \left. G_{\bf r}({\bm \eta}) \right|_{\eta_k=1}
    .
    \label{}
\end{equation}
In this way, a click event in the output mode $k$, corresponding to the detection of one or more photons, is defined as the complement of the vacuum event and is described by the POVM element
\begin{equation}
    \hat{\Pi}_{l_k} = \hat{I}_k - \hat{\Pi}_{0_k} ,
    \label{POVM_click}
\end{equation}
where $\hat{I}_k$ is the identity operator in the corresponding mode $k$. Note that, from Eq.~(\ref{gen_func_smss}), we can readily identify how the
identity operator $\hat{I}_k$ acts on each mode via the property
$\left. G_{\bf r}({\bm \eta}) \right|_{{\bm \eta} = {\bm 0}} = 1$.
Therefore, it follows that the probability of a click event is given by
\begin{equation}
    P_{\bf r}(l_k) \equiv \langle \hat{\Pi}_{l_k}  \rangle = \left. G_{\bf r}({\bm \eta}) \right|_{\eta_k=0} -  \left. G_{\bf r}({\bm \eta}) \right|_{\eta_k=1}.
    \label{probability_click}
\end{equation}

We now consider the event in which $N$ clicks are registered in the set of output modes $L\equiv\{k_1,\ldots,k_N\}$, where each $k_j\in\{1,\ldots,M\}$ labels a distinct mode in which a click is detected, while in the remaining output modes no photons are detected. We denote such a configuration by ${\bf l}=(l_{k_1},\ldots,l_{k_N})$.  In this way, it is convenient to introduce the reduced generating function $G_{\bf r}({\bm \eta}')$, defined by setting $\eta'_k=\eta_k$ if $k \in L$ and
$\eta'_k=1$ if $k \notin L$, where the latter accounts for the vacuum detections according to Eq.~(\ref{probability_no_click}). Therefore, the probability of detecting the corresponding click-event configuration ${\bf l}$ follows directly from Eqs.~(\ref{POVM_click}) and~(\ref{probability_click}) as
\begin{align}
    P_{\bf r}({\bf l}) & \equiv \langle \prod^M_{k=1} \hat{\Pi}_{l_k} \rangle 
    \nonumber\\
    &= \sum_{S  \subseteq L} (-1)^{|S|} \, \langle  \prod_{k \in S} \hat{\Pi}_{0_k} \prod_{k \notin S} \hat{I}_k  \rangle 
    \nonumber\\
    &= \sum_{S  \subseteq L} (-1)^{|S|} \, G_{\bf r}({\bm \eta}') \Big|_{\substack{\eta'_k = 1, k \in S \\ \eta'_k = 0, k  \notin S}}
    \label{probability_click}
    ,
\end{align}
where the sum over $S\subseteq L$ runs over all subsets of $L$, that is, over all possible choices of modes selected from $L$. Therefore, the equation~(\ref{probability_click}) extends the result of Ref.~\cite{QuesadaGBSClick} beyond the fully indistinguishable case, providing an explicit expression for click probabilities in the presence of partial distinguishability.


\section{\label{sec:application1} Internal states with homogeneous overlap}

In this section, we specialize our general framework to the homogeneous distinguishability model, which provides a useful way to investigate partial distinguishability as an effective noise source in the boson sampling schemes~\cite{BSAlg1, Shi2022, GBSBinnedValidation}. This enables us to
investigate how our formalism operates within this model and cleanly isolate the contribution of distinguishability to the output statistics. Thus, the partial distinguishability is modeled by decomposing the internal state of each photon into a common (indistinguishable) mode $|\phi_0\rangle$ and an orthogonal mode $|\phi_k^\perp\rangle$, 
\begin{equation}
    | \psi_k \rangle = \sqrt{1-\epsilon} \, | \phi_0 \rangle + \sqrt{\epsilon} \, | \phi^\perp_k \rangle ,
    \label{model_homogeneous_dist}
\end{equation}
where $0 \leq \epsilon \leq 1$ quantifies the degree of distinguishability, with $ \langle \phi_0 | \phi^\perp_j \rangle = 0$ and $ \langle \phi^\perp_k | \phi^\perp_j \rangle = \delta_{kj}$. In this decomposition, the component $|\phi_0\rangle$ results in the genuine multiphoton interference, whereas the orthogonal components $|\phi_k^\perp\rangle$ contribute without interference and
are therefore interpreted as the classical contribution. Thus, the overlap matrix V, defined in Eq.~(\ref{def_H}), reduces to
\begin{equation}
    V_{ij} = (1-\epsilon) + \epsilon \, \delta_{ij} ,
    \label{v_m_model}
\end{equation}
where $I_M$ denotes the usual $M \times M$ identity matrix. Inserting Eq.~(\ref{v_m_model}) into Eqs.~(\ref{def_H}) and (\ref{H_r}), we obtain the matrix $\mathcal{M}_{\bf r}$ as a convex sum of the two extremal cases, as follows:
\begin{equation}
    \mathcal{M}_{\bf r} = (1-\epsilon) \, \mathcal{M}^0_{\bf r} + \epsilon \, \mathcal{M}^\perp_{\bf r},
    \label{H_r_homogeneous}
\end{equation}
where the first term corresponds to the indistinguishable contribution introduced in Eq.~(\ref{H_r_ind}), while the second term defines the distinguishable (orthogonal) contribution:
\begin{equation}
    \mathcal{M}^\perp_{\bf r}=
    \begin{pmatrix}
    0_M & H^\perp_{\bf r}\\
    \overline{H}^\perp_{\bf r} & 0_M
    \end{pmatrix}
    ,
    \quad
    H^\perp_{\bf r} \equiv \big. H_{\bf r} \big|_{V_{ij}=\delta_{ij}}
    ,
    \label{H_r_dist}
\end{equation}
which follows that $H^\perp_{\bf r} = \text{diag}(H^0_{\bf r})$. Thus, substituting Eq.~(\ref{H_r_homogeneous}) into Eq.~(\ref{gen_func_smss}), we obtain the corresponding expression for the generating function
\begin{equation}
    G_{\bf r}({\bm \eta}, \epsilon) = c_{\bf r} \,
    \text{det}\left[I_{2M} - (1-\epsilon) \, \mathcal{M}^0_{\bf r} - \epsilon \, \mathcal{M}^\perp_{\bf r} \right]^{-\frac{1}{2}} 
    .
    \label{gen_func_homog}
\end{equation}

Note that, when the generating function factorizes as
$G_{\bf r}({\bm \eta},\epsilon)=g_1({\bm \eta})\,g_2({\bm \eta})$, it follows that the probability can be expressed as a sum of two contributions. Using the compact notation $\partial_{\eta_l}\equiv\partial/\partial\eta_l$, this property is proven by applying the generalized Leibniz rule to the factorized generating function, 
\begin{align}
    &(-\partial_{\eta_l})^{n_l}
    \, g_1({\bm \eta})\, g_2({\bm \eta})
    \nonumber\\
    &= \sum_{m_l=0}^{n_l} \binom{n_l}{m_l}
    \Big[(-\partial_{\eta_l})^{m_l} g_1({\bm \eta})\Big]
    \Big[(-\partial_{\eta_l})^{n_l-m_l} g_2({\bm \eta})\Big] 
    \label{Leibniz_rule}
    ,
\end{align}
and then substituting in Eq.~(\ref{probability_gen_func}).
Motivated by this observation, we present a convenient factorization of the generating function in Eq.~(\ref{gen_func_homog}), separating the orthogonal component defined in Eq.~(\ref{H_r_dist}), resulting in
\begin{align}
    G_{\bf r}({\bm \eta}, \epsilon) 
    &= c_{\bf r} \, 
    \text{det}(I_{2M} - \epsilon \, \mathcal{M}^\perp_{\bf r})^{-\frac{1}{2}} 
    \nonumber\\
    & \quad \times
    \text{det}\left[I_{2M} - (1-\epsilon) \, \widetilde{\mathcal{M}}^0_{\bf r} \right]^{-\frac{1}{2}}
    ,
    \label{gen_func_factor}
\end{align}
where we define the noise-modified version of the matrix  $\mathcal{M}^0_{\bf r}$ defined in Eq.~(\ref{H_r_ind}), as follows
\begin{equation} 
    \widetilde{\mathcal{M}}^0_{\bf r}  = 
    (I_{2M} - \epsilon \, \mathcal{M}^\perp_{\bf r})^{-1}
    \mathcal{M}^0_{\bf r}
    .
    \label{tilde_matrix_M}
\end{equation}
Thus, substituting  Eq.~(\ref{gen_func_factor}) into Eq.~(\ref{probability_gen_func}) and applying Eq.~(\ref{Leibniz_rule}), the output probability can be written as a sum of a classical contribution $P^{\perp}_{\mathbf r}(\mathbf m)$ and a noisy quantum contribution $\widetilde{P}^{0}_{\mathbf r}(\mathbf n-\mathbf m)$,
\begin{equation}
    P_{\bf r}({\bf n}) = c_{\bf r}  \sum^{n_1}_{m_1=0} ... \sum^{n_M}_{m_M=0} \epsilon^{|{\bf m }|} \, 
    P^\perp_{\bf r}({\bf m }) \, 
    \widetilde{P}^0_{\bf r}({\bf n}-{\bf m }) 
    .
    \label{prob_dist_indist}
\end{equation}

We now focus on the explicit evaluation of the classical contribution in Eq.~(\ref{prob_dist_indist}). From Eq.~(\ref{H_r_dist}), we have
\begin{align}
    &\text{det}
    (I_{2M} - \epsilon \, \mathcal{M}^\perp_{\bf r}) 
    \nonumber\\
    &= \prod^M_{k=1} \left[ 1 - \epsilon^2 \tanh^2 r_k \left( \sum^M_{l=1} |U_{kl}|^2 (1-\eta_l) \right)^2 \right]
    .
    \label{det_matrix_N}
\end{align}
Recalling that squeezing is applied only to the first $N$ modes, so that $\tanh r_k=\tanh r$ for $1\le k\le N$ and $\tanh r_k=0$ for $N<k\le M$, and additionally assuming a uniform multiport $|U_{kl}|^2=1/M$, the classical probability
in Eq.~(\ref{prob_dist_indist}) follows directly from
Eqs.~(\ref{det_matrix_N}) and (\ref{probability_gen_func}) as
\begin{align}
    P^\perp_{\bf r} ({\bf m})
    &=  \left. \prod^M_{k=1} \frac{\eta^{m_k}_k}{m_k!} \left( - \frac{\partial}{\partial \eta_k} \right)^{m_k} \text{det}
    (I_{2M} - \mathcal{M}^\perp_{\bf r})^{-\frac{1}{2}} \right|_{{\bm \eta} = {\bm 1}} 
    \nonumber\\
    & = 
    \begin{cases}
    \frac{2m!}{m!} \left(\frac{N}{2} \right)_m \left( \frac{\tanh \, r}{M}\right)^{2m} & , |{\bf m}| = 2 m \\
    0 & , |{\bf m}| = 2 m+1.
    \end{cases}
    \label{dist_probability}
\end{align}
where $(N/2)_m=\prod_{r=0}^{m-1}(N/2+r)$ denotes the rising factorial. The same final expression for $P^\perp_{\bf r} ({\bf m})$ in Eq.~(\ref{dist_probability}) was also reported in Ref.~\cite{ValeryGBS}, obtained by a first-quantization method. The remaining term $\widetilde{P}^{0}_{\bf r}({\bf n})$ can be formally obtained by substituting the generating-function factor containing $\widetilde{\mathcal{M}}^0_{\bf r}$ into Eq.~(\ref{probability_gen_func}). In the indistinguishable limit ($\epsilon=0$), it reduces to the ideal probability $P^{0}_{\bf r}({\bf n})$ defined in Eq.~(\ref{probability_hafnian}).

An important observation is that the distinguishable component in Eq.~(\ref{gen_func_factor}) gives
the classical probabilities $P^\perp_{\bf r}({\bf n})$, which are easily computed, as shown in Eq.~(\ref{dist_probability}). Thus, the remaining term $\widetilde{P}^0_{\bf r}({\bf n})$ contains all the genuine
multiphoton-interference effects and therefore encodes the quantum contribution responsible for computational hardness. Such a decomposition of the probability into a classical (distinguishable)
part and a quantum (interfering) part was also identified in FBS~\cite{ValeryPRA2019}.

\subsection{Distinguishability contribution as \\ an average over displaced states}

In the large noise regime ($\epsilon \approx 1$), the distinguishability arising from orthogonal internal modes suppresses the multiphoton interference, collapsing the probability into Eq.~(\ref{dist_probability}). We therefore focus now on the low-noise regime ($\epsilon \approx 0$), to elucidate how the partial distinguishability affects the ideal squeezed state interference. In this regime, we identify an additional interpretation in which the contribution from the orthogonal modes can be understood as an average over displaced states.

We start by expressing the input creation operators in the exponent of Eq.~(\ref{smss_input}) explicitly in the internal-state basis defined in Eq.~(\ref{model_homogeneous_dist}), resulting in
\begin{align}
    &| {\bf r} (\epsilon) \rangle
    \nonumber\\
    &= c^{\frac{1}{2}}_{\bf r} \, \prod_{k=1}^M \exp 
    \left[ \frac{S_k}{2}
    \left( \sqrt{1-\epsilon} \, \hat{a}^\dagger_{k, \phi_0}
    + \sqrt{\epsilon} \, \hat{a}^\dagger_{k, \phi^\perp_k} \right)^2
    \right]  |0 \rangle 
    \nonumber\\
    &= c^{\frac{1}{2}}_{\bf r} \, \prod_{k=1}^M \exp \left[
    \frac{S_k}{2}(\hat{a}^\dagger_{k, \phi_0})^2  + 
    S_k \sqrt{\epsilon} \, \hat{a}^\dagger_{k, \phi_0} \hat{a}^\dagger_{k, \phi^\perp_k} + \mathcal{O}(\epsilon) 
    \right] |0 \rangle
    .
    \label{smss_homog}
\end{align}
Note that, in this approximation, we neglect second-order contributions from the orthogonal modes $(\hat{a}^\dagger_{k,\phi_k^\perp})^2$, which effectively produce the trivial probabilities given by
Eq.~(\ref{dist_probability}).

In sequence, we decompose the photon-counting operator by separating the detection of the common mode $|\phi_0\rangle$, and the detection of the orthogonal modes $|\phi_k^\perp\rangle$, in Eq.~(\ref{photon_number_def}), resulting:
\begin{equation}
    \hat{N}_l = \eta_l \, \hat{b}^\dagger_{l,\phi_0} \hat{b}_{l,\phi_0} + \eta_l \sum^M_{k=1} \hat{b}^\dagger_{l,\phi^\perp_k} \hat{b}_{l,\phi^\perp_k} .
    \label{detec_oper_homog}
\end{equation}
Using Eqs.~(\ref{vacuum_prob}) and (\ref{gen_func_definition}), the generating function can be rewritten in a form where the detection acts explicitly only on the common mode, while the all the contributions from the orthogonal components are absorbed into an effective Husimi function that retains all the noise effects,
\begin{align}
    G_{\bf r}({\bm \eta}, \epsilon) &= \int d^2 {\bm \alpha}_{\phi_0} \, Q_{\text{eff}}({\bm \alpha}_{\phi_0}) \nonumber\\
    &\quad \times \, \frac{\text{exp}\left[- {\bm \alpha}^\dagger_{\phi_0} \left(H^{-1}_0 - I_M \right) {\bm \alpha}_{\phi_0} \right]}{\text{det}(H_0)} ,
    \label{gen_func_aux2}
\end{align}
where from Eq.~(\ref{def_H}) we define $H_0 \equiv U \Lambda U^\dagger$. In Eq.~\ref{gen_func_aux2} we define the phase-space coordinates that parameterize the Husimi function statistics of the common mode as ${\bm \alpha}_{\phi_0} = (\alpha_{1,\phi_0}, ..., \alpha_{M,\phi_0})$. The detailed derivation is presented in Appendix.~\ref{detailed_aver_coherent}

Under the approximation in Eq.~(\ref{smss_homog}), the effective Husimi function $Q_{\mathrm{eff}}({\bm \alpha}_{\phi_0})$ in Eq.~(\ref{gen_func_aux2}) takes the form
\begin{align}
    Q_{\text{eff}}({\bm \alpha}_{\phi_0}) &\approx \frac{c_{\bf r}}{\pi^M} \, \langle {\bm \alpha}_{\phi_0} |
    \prod_{k=1}^M \exp \left[
    \frac{S_k}{2}(\hat{a}^\dagger_{k, \phi_0})^2 \right]
    \,
    \nonumber\\
    & \quad \times \hat{\rho}_\epsilon  \, \prod_{k=1}^M \exp \left[ \frac{S^*_k}{2}\hat{a}_{k, \phi_0}^2 \right]
    | {\bm \alpha}_{\phi_0} \rangle
    \label{Husimi_effective}
    ,
\end{align}
where $\hat{\rho}_\epsilon$ collects all noise contributions in the model. In the following, we introduce the phase-space coordinates of the orthogonal modes as ${\bm \alpha}_{\phi_\perp}=(\alpha_{1,\phi^\perp_1},\ldots,\alpha_{M,\phi^\perp_M})$. Therfore,  the state $\hat{\rho}_\epsilon$ in Eq.~(\ref{Husimi_effective}) is given as a mixture of displaced states, in which the orthogonal modes enter as displacement parameters, weighted by a Gaussian distribution with diagonal covariance matrix $\Sigma=\epsilon D_{\bf r}\,\mathrm{diag}(H_0)\,\overline{D}_{\bf r}$,
\begin{align}
    \hat{\rho}_\epsilon &=  \int \frac{d^2 {\bm \beta}}{\pi^M} \, \frac{\text{exp}\left(- {\bm \beta}^\dagger \Sigma^{-1} {\bm \beta} \right)}{\text{det}(\Sigma)} 
    \nonumber\\
    & \quad \times \text{exp}\left( {\bm \beta}^t {\bf \hat{a}}^\dagger_{\phi_0}\right) | 0 \rangle \langle 0 | \,\text{exp}\left( {\bm \beta}^\dagger {\bf \hat{a}}_{\phi_0}\right)
    .
    \label{rho_effective}
\end{align}
where we introduce the effective displacement
${\bm \beta}=\sqrt{\epsilon}\,\overline{D}_{\bf r}\,{\bm \alpha}_{\phi_\perp}$, the creation-operator vector on the common mode ${\bf \hat{a}}^\dagger_{\phi_0}=(\hat{a}^\dagger_{1,\phi_0},\ldots,\hat{a}^\dagger_{M,\phi_0})$,
and define ${\bf \hat{a}}_{\phi_0}$ analogously. In the noiseless limit, Eq.~(\ref{rho_effective})  becomes the vacuum,
\begin{equation}
    \big. \hat{\rho}_\epsilon \big|_{\epsilon=0} =|0\rangle \langle 0|
    ,
\end{equation}
and therefore the effective Husimi function in Eq.~(\ref{Husimi_effective}) reduces to that of (vacuum) squeezed-states, as expected. Detailed derivations of Eqs.~(\ref{gen_func_aux2}), (\ref{Husimi_effective}), and
(\ref{rho_effective}) are given in Appendix~\ref{detailed_aver_coherent}.

Our approach also recovers a known result for the effect of losses on squeezed states, first derived in Ref.~\cite{PatronSimulabilityGBS}, where a homogeneous loss maps a pure squeezed state to a mixed Gaussian state, equivalently to a
squeezed thermal state. In our framework, this corresponds to the case where photons occupying orthogonal modes are not detected. Thus, this result is recovered by setting ${\bm \eta}={\bm 0}$ in Eq.~(\ref{rho_effective}), which is equivalent to taking $\eta_l=0$ for all $l$ in the detection operator for the orthogonal modes in Eq.~(\ref{detec_oper_homog}). In this way, Eq.~(\ref{rho_effective}) reduces to a thermal state in the
$P$ representation, with the following parametrization:
\begin{equation}
    \left. \Sigma_{kk} \right|_{\eta_k=0} = \epsilon  \tanh^2 r_k = \frac{\overline{n}_k+1}{\overline{n}_k},
    \label{aver_thermal_photons}
\end{equation}
where $\overline{n}_k$ denotes the average number of thermal photons. In addition, $\overline{n}_k$ can be written as $\overline{n}_k = \sinh^2 \widetilde{r}_k$, with $\tanh \widetilde{r}_k = \sqrt{\epsilon}\,\tanh r_k$.


\section{\label{sec:application2} Gaussian model for the internal states and zero-probability events}

In this section, we address a more fundamental aspect of partial distinguishability. In two-photon interference, described by the HOM effect~\cite{HOM1987}, the distinguishability is fully captured by the squared modulus of the mutual overlap, $|V_{12}|^{2}$ \cite{HOM1987,Loudon1989,Mandel1991}, often referred to as the visibility. By contrast, for interferences of three or more photons in Fock states, partial distinguishability depends not only on the moduli of pairwise overlaps but also on phases encoded in the internal states. In particular, three-photon interference is sensitive to a collective three-particle phase \cite{CollectivePhase1} given by $\arg\!\left(V_{12}V_{23}V_{31}\right)$, which is not determined by the pairwise moduli $|V_{ij}|^{2}$ alone. Likewise, four-photon interference exhibits an analogous four-particle phase \cite{CollectivePhase3} and, more generally, collective phases associated with genuine multiphoton overlaps \cite{CollectivePhase1}. To the best of our knowledge, such phase-dependent effects have not yet been investigated in the interference of single-mode squeezed states.

For this purpose, we adopt a realistic Gaussian model for the internal state of the photons, which has been widely employed in previous studies~\cite{ValeryBS1,ValeryBS2,CollectivePhase1,CollectivePhase2,CollectivePhase3} because it provides a clear physical interpretation and an experimentally motivated description of the internal state overlap. In this way, the internal state of the photon prepared in the input mode $k$ is defined in a temporal-mode basis as
\begin{equation}
    | \psi_k \rangle = \int d t \, \psi_k(t-T_k) | t \rangle
    ,
    \label{smss_gauss}
\end{equation}
where we define the temporal wave-packet as the following Gaussian distribution
\begin{equation}
    \psi_k(t) = \frac{1}{\left(\pi \sigma^2_t \right)^{\frac{1}{4}}} \exp \left[ - \left(\frac{t}{2 \sigma_t} \right)^2 + i \Omega_0 t \right] 
    ,
\end{equation}
parameterized by the relative time delay $T_k$, central frequency $\Omega_0$, and pulse width $\sigma_t$, resulting in the overlap
\begin{equation}
    V_{kj} = \exp \left[ - \left(\frac{T_k-T_j}{2 \sigma_t} \right)^2 + i \, \Omega_0 (T_k-T_j) \right] 
    .
    \label{smss_gauss_product}
\end{equation}

In the following, we first derive the zero-output probabilities for perfectly indistinguishable photons and sources with identical squeezing parameters,
and then investigate how these events are modified by partial
distinguishability.

To begin, we consider interference at a balanced beamsplitter with a reflection phase of $\pi/2$, which implements the mode transformation defined in
Eq.~(\ref{spatial_transf}) via the unitary matrix
\begin{equation}
    U^{(2)} = \frac{1}{\sqrt{2}}
    \begin{pmatrix}
        1 & i \\
        i & 1
    \end{pmatrix} 
    .
    \label{matrix_beamsplitter}
\end{equation}
For perfectly indistinguishable photons, all modes share the same internal state, $|\psi_k \rangle=|\phi_0 \rangle$ in Eq.~(\ref{smss_input}). Applying the state evolution
performed by Eq.~(\ref{matrix_beamsplitter}) to the creation operators in the exponent of Eq.~(\ref{smss_input}), we obtain the following output state
\begin{align}
    | r_1, r_2 \rangle_{(out)} 
    &= c^{\frac{1}{2}}_{\bf r} \, \text{exp} \left[ \frac{S}{2} \sum^2_{k=1}\left( \sum^2_{l=1} U^{(2)}_{kl} \hat{b}^\dagger_{l, \phi_0} \right)^2 \right] |0 \rangle
    \nonumber\\
    &= c^{\frac{1}{2}}_{\bf r} \exp \left( i \, S \, \hat{b}^\dagger_{1,\phi_0} \hat{b}^\dagger_{2,\phi_0}  \right) |0 \rangle 
    ,
\end{align}
where we assumed $S_1=S_2=S$. Therefore, the output state $| r_1, r_2 \rangle_{(out)}$ is effectively a two-mode squeezed state, implying that only outcomes with equal photon numbers in the two output modes are allowed, resulting in $P_{\bf r}(n_1, n_2)=0$ for any $n_1\neq n_2$. 

\begin{figure}[t]
    \centering
\includegraphics[width=0.9 \columnwidth]{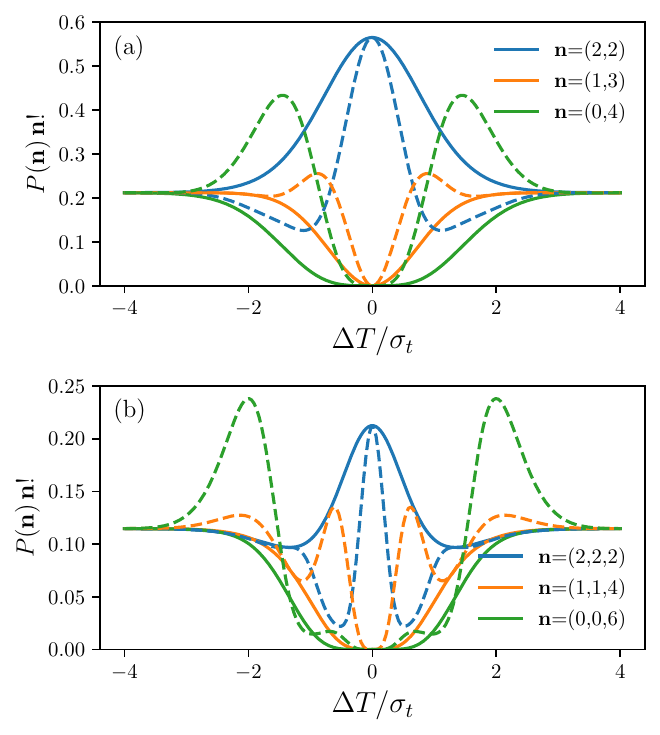}
\caption{Output probabilities for the interference of single-mode squeezed states with squeezing parameters $r_k=1$ in (a) a balanced beamsplitter and (b) a balanced tritter. The probabilities are shown as a function of the normalized relative delay $\Delta T/\sigma_t$ for two central frequencies, $\Omega_0=0$ (solid) and $\Omega_0=\pi/4$ (dashed).}
\label{probabilities_smss}
\end{figure}

To investigate the effects of partial distinguishability on these probabilities, Fig.~\ref{probabilities_smss}.(a) shows the selected output probabilities $P_{\bf r}(n_1,n_2)$ as a function of the relative time delay $\Delta T=T_1-T_2$, for two different central frequencies $\Omega_0$. As in the original HOM effect, the destructive interference is maximal at $\Delta T=0$ and continuously degraded as $|\Delta T|$ increases. Additionally, Fig.~\ref{probabilities_smss}.(a) reveals a dependence on the complex phase of $V_{12}$, arising by the difference choices of the central frequency $\Omega_0$ in the dashed and solid curves. In contrast to the two-photon Fock-state case, for squeezed-state inputs, this phase dependence arises already at the two-mode level because the input is a coherent superposition of photon-pair components rather than a state with definite photon number.

In sequence, we consider interference at a balanced tritter, which implements the mode transformation defined in
Eq.~(\ref{spatial_transf}) via the unitary matrix
\begin{equation}
    U^{(3)} = \frac{1}{\sqrt{3}}
    \begin{pmatrix}
        1 & 1 & 1 \\
        1 & \text{e}^{i\frac{2\pi}{3}} & \text{e}^{-i\frac{2\pi}{3}} \\
        1 & \text{e}^{-i\frac{2\pi}{3}} & \text{e}^{i\frac{2\pi}{3}}
    \end{pmatrix} .
    \label{matrix_tritter}
\end{equation}
Analogously to the two-mode case, substituting the transformation in Eq.~(\ref{matrix_tritter}) into Eq.~(\ref{smss_input}) for perfectly indistinguishable photons gives the output state
\begin{align}
    &| r_1, r_2, r_3 \rangle_{(out)} 
    \nonumber\\
    &=c^{\frac{1}{2}}_{\bf r} \, \text{exp} \left[ \frac{S}{2} \sum^3_{k=1}\left( \sum^3_{l=1} U^{(3)}_{kl} \hat{b}^\dagger_{l, \phi_0} \right)^2 \right] |0 \rangle
    \nonumber\\
    &= c^{\frac{1}{2}}_{\bf r}  \exp \left[ S \left( (\hat{b}^\dagger_{1,\phi_0})^2 + 2 \, \hat{b}^\dagger_{2,\phi_0} \hat{b}^\dagger_{3,\phi_0} \right) \right] |0 \rangle ,
\end{align}
where $S_1=S_2=S_3$. Therefore, the output state $| r_1, r_2, r_3 \rangle_{(out)} $ factorizes into an effective single-mode squeezed state in mode $l=1$ and a two-mode squeezed state shared by modes $l=2,3$. It follows that $P_{\bf r}(n_1,n_2,n_3)=0$ whenever $n_1$ is odd or $n_2\neq n_3$. 

Finally, to investigate the effects of partial distinguishability on these
probabilities, Fig.~\ref{probabilities_smss}(b) shows selected output probabilities $P_{\mathbf r}(n_1,n_2,n_3)$ for equal temporal delays between neighboring inputs, so that $\Delta T \equiv T_1-T_2 = T_2-T_3$ and $T_1-T_3 = 2\Delta T$. As expected, interference is maximal at $\Delta T=0$ and gradually degrades
as $|\Delta T|$ increases. The probabilities also exhibit a pronounced dependence on the overlap phases, controlled by the choices of $\Omega_0$.


\section{Conclusion}

In this work, we apply a generating-function framework based on the phase-space formalism to describe the interference of squeezed-states in linear interferometers with partially distinguishable photons. This approach gives a general expression for output, where the distinguishability enters explicitly through the overlap matrix appearing in the generating function. By applying the well-known homogeneous overlap model in our results, we obtain a clear interpretation of how distinguishability manifests in the output probabilities, which decompose into a classical contribution that progressively suppresses multiphoton interference and degrades the quantum advantage. These results provide a unified and physically transparent description of distinguishability effects in squeezed-state interferometry and offer a useful tool for assessing the impact of realistic imperfections in photonic quantum-information experiments. From a more fundamental perspective, we also provide a experimentally motivated route to probe partial distinguishability effects beyond visibility by explicitly investigating the phases carried by the internal states of the photons.


\begin{acknowledgments}

M. E. O. B. thanks the São Paulo Research Foundation (FAPESP), Process Number 21/03251-0, and the Coordenação de Aperfeiçoamento de Pessoal de Nível Superior (CAPES) for the financial support. V.S. thanks the National Council for Scientific and Technological Development of Brazil (CNPq) for the financial support.

\end{acknowledgments}


\newpage
\onecolumngrid
\appendix

\section{\label{appendix_gen_func_smss} Derivation of the squeezed-state generating function}

In this section, we recall some important results derived in Ref.~\cite{Valery2017} and following that, provide a detailed calculation of the generating function for the squeezed state interference, shown in Eq.~(\ref{gen_func_smss}). Introducing the annihilation-operator vector
${\bf \hat{a}}_{\Psi} = (\hat{a}_{1,\psi_1}, \ldots, \hat{a}_{M,\psi_M})$,
and summing the photon-number operators defined in Eq.~(\ref{photon_number_def}) one finds \cite{Valery2017}
\begin{equation}
    \sum^M_{l=1} \hat{N}_l 
    =\sum^M_{l=1} \eta_l \sum^D_{s=1} \sum^M_{k_1=1} \sum^M_{k_2=1} U_{l,k_1} \overline{U}_{l,k_2} \, \overline{\phi}_{k1,s} \phi_{ks} \hat{a}^\dagger_{k_1, \psi_{k_1}} \hat{a}_{k_2, \psi_{k_2}} 
    = {\bf \hat{a}}^\dagger_{\Psi} \left(I_M - H \right) {\bf \hat{a}}_{\Psi}
    ,
    \label{summation_Nl}
\end{equation}
after inverting the transformations implemented by Eqs.~(\ref{spatial_transf}) and~(\ref{internal_transf}) and defining the matrix $H$ as in Eq.~(\ref{def_H}). This representation allows one to convert the normal ordered operator appearing in Eq.~(\ref{vacuum_prob}) into the anti-normal ordered form using the identity \cite{Valery2017}
\begin{equation}
    \langle \, \mathcal{N} \big\{ \exp \left[-\hat{\bf a}^\dagger (I_M-H)\hat{\bf a}_{\Psi} \right] \big\} \, \rangle
    = \frac{1}{\det(H)}
    \langle \, \mathcal{A} \big\{ \exp \left[-\hat{\bf a}^\dagger_{\Psi} (H^{-1}-I_M)\hat{\bf a}_{\Psi} \right] \big\} \, \rangle ,
    \label{normal_antinormal}
\end{equation}
and substituting into Eq.~(\ref{vacuum_prob}), we obtain
\begin{equation}
    G({\bm \eta}) =  \frac{1}{\det(H)}
    \langle \, \mathcal{A} \big\{ \exp \left[-\hat{\bf a}^\dagger_{\Psi} (H^{-1}-I_M) \hat{\bf a}_{\Psi} \right] \big\} \, \rangle .
    \label{vacuum_prob_aux}
\end{equation}
Finally, using the quantum equivalence theorem, the expectation value in Eq.~(\ref{vacuum_prob_aux}) can be expressed in terms of the Husimi $Q$ function, which directly results in Eq.~(\ref{gen_func_definition}).

Next, introducing the phase-space coordinates ${\bm \alpha} = (\alpha_1, \ldots, \alpha_M)$, we define a coherent state with photons occupying arbitrary internal states in each input mode $k$ as
\begin{eqnarray}
    | \alpha_k  \rangle = \text{exp} \Big( - |\alpha_k|^2 + \alpha_k \hat{a}^\dagger_{k, \psi_k} \Big) |0 \rangle 
    ,
    \label{co_input}
\end{eqnarray}
From the definition of the squeezed states given in Eq.~(\ref{smss_input}), we obtain the corresponding Husimi function,
\begin{equation}
    Q_{\bf r}({\bm \alpha}) = \prod^M_{k=1} \frac{1}{\pi} |\langle \alpha_k  | r_k \rangle|^2 ,
    \label{hussimi_smss_def}
\end{equation}
with the overlap explicitly given by
\begin{equation}
    \langle \alpha_k | r_k \rangle = \sqrt{1-|S_k|^2} \, \text{exp} \left[-|\alpha_k|^2 - \frac{1}{2} \left( S_k \alpha^2_k  + c.c. \right) \right] ,
    \label{overlap_coherent_smss}
\end{equation}
with  $S_k = \text{tanh} \, r_k \, \text{e}^{i \theta_k}$. Introducing the complete phase-space coordinate vector
$\mathcal{A} = ({\bm \alpha}, {\bm \alpha}^*)$,
one may identify the identity
\begin{equation}
    {\bm \alpha}^\dagger X {\bm \alpha} =
    \frac{1}{2} \mathcal{A}^\dagger 
    \begin{pmatrix}
        X & 0_M \\
        0_M & \overline{X}
    \end{pmatrix}
    \mathcal{A},
    \label{identity_phasespace_coordinates}
\end{equation}
which holds for any $M\times M$ Hermitian matrix $X$. As a result, the Husimi function of the squeezed state given in Eq.~(\ref{hussimi_smss}) follows directly by substituting Eq.~(\ref{overlap_coherent_smss}) into Eq.~(\ref{hussimi_smss_def}) and rewriting the exponent in terms of the diagonal matrix
$D_{\bm r} = \mathrm{diag}(S_1,\ldots,S_M)$ and the identity matrix $I_M$.

Next, the generating function of the squeezed state is evaluated from Eq.~(\ref{gen_func_definition}). Substituting the Husimi function $Q_{\bf r}({\bm \alpha})$ obtained in Eq.~(\ref{hussimi_smss}) and rewriting the exponent using the identity in Eq.~(\ref{identity_phasespace_coordinates}), we arrive at the integral expression given in Eq.~(\ref{gen_func_smss_integral}). That integral can be evaluated using the relation
\begin{equation}
    \int \frac{d^2{\bm \alpha}}{\pi^M} ~ \text{exp}\left( - \frac{1}{2} \mathcal{A}^\dagger Z \mathcal{A} \right) = \frac{1}{\sqrt{\text{det}(Z)}}
    \label{det_z_1}
    ,
\end{equation}
where $Z$ is readily identified from Eq.~(\ref{gen_func_smss_integral}), resulting in the following determinant after some straightforward algebra
\begin{equation}
    \text{det}(Z) 
    = 
    \text{det}
    \begin{pmatrix}
        H^{-1} & \overline{D}_{\bf r} \\
        D_{\bf r} & \overline{H}^{-1}
    \end{pmatrix} 
    = \frac{1}{\text{det}(H)} \, \text{det}\left( H^{-1} - \overline{D}_{\bf r} \overline{H} D_{\bf r} \right) 
    = \frac{1}{\text{det}(H)^2} \, \text{det}\left( I_M - H_{\bf r} \overline{H}_{\bf r} \right)
    .
    \label{det_z_2}
\end{equation}
Finally, substituting Eqs.~(\ref{det_z_1}) and~(\ref{det_z_2}) in (\ref{gen_func_smss_integral}) we arrive at the final expression for the generating function for the squeezed state presented in Eq.~(\ref{gen_func_smss}) at the main text.

\section{\label{detailed_aver_coherent} Detailed calculations for the \\ distinguishability  as an average over displacements}

In this section we explain in details the derivation of Eqs.~(\ref{gen_func_aux2}), (\ref{Husimi_effective}), and
(\ref{rho_effective}) presented in the main text. To begin, it is convenient to rewrite Eq.~(\ref{gen_func_definition}) in terms of a Gaussian kernel $\mathcal{K}({\bm \alpha},H)$ as follows:
\begin{align}
    G({\bm \eta}) = \int d^2{\bm \alpha} \, Q({\bm \alpha}) \, \mathcal{K}({\bm \alpha},H) 
    ,
    \quad
    \mathcal{K}({\bm \alpha},H) = \frac{\text{exp}\left[- {\bm \alpha}^\dagger \left(H^{-1} - I_M \right) {\bm \alpha} \right]}{\text{det}(H)} .
    \label{kernel_def}
\end{align}

In the following, we analyze the decomposition of the photon-counting operator introduced in Eq.~(\ref{detec_oper_homog}). Since the operator $\hat{N}_l$ has already been explicitly expanded in the internal basis, we may directly apply the relation given in Eq.~(\ref{summation_Nl}) by setting $V_{ij}=1$, which results
\begin{align}
    \sum^M_{l=1} \hat{N}_l = {\bf \hat{a}}^\dagger_{\phi_0} \left(I_M - H_0 \right) {\bf \hat{a}}_{\phi_0}  + {\bf \hat{a}}^\dagger_{\phi_\perp} \left(I_M - H_\perp \right) {\bf \hat{a}}_{\phi_\perp} 
    \label{summation_Nl_model}
\end{align}
with $H_0 = U \Lambda U^\dagger$ and $H_\perp = \mathrm{diag}(H_0)$. In Eq.~(\ref{summation_Nl_model}), we introduce the annihilation-operator vectors associated with the common mode and the orthogonal modes, respectively, as ${\bf \hat{a}}_{\phi_0} = (\hat{a}_{1,\phi_0}, \ldots, \hat{a}_{M,\phi_0})$ 
and ${\bf \hat{a}}_{\phi_\perp} = (\hat{a}_{1,\phi^\perp_1}, \ldots, \hat{a}_{M,\phi^\perp_M})$. Substituting Eq.~(\ref{summation_Nl_model}) into the definition of the generating function given in Eq.~(\ref{vacuum_prob}), and making use of the normal-antinormal ordering identity in Eq.~(\ref{normal_antinormal}), we obtain
\begin{align}
    G_{\bf r}(\bm{\eta}, \epsilon)
    &= \left\langle 
    \mathcal{N} \left\{ 
    \exp \left[ {\bf \hat{a}}^\dagger_{\phi_0} \left(I_M - H_0 \right) {\bf \hat{a}}_{\phi_0} \right\} 
     \mathcal{N} \left\{  {\bf \hat{a}}^\dagger_{\phi_\perp} \left(I_M - H_\perp \right) {\bf \hat{a}}_{\phi_\perp}  \right]
    \right\} \right\rangle  
    \nonumber\\
    &= \left\langle 
    \frac{\mathcal{A} \left\{\exp \left[ {\bf \hat{a}}^\dagger_{\phi_0} \left(I_M - H_0 \right) {\bf \hat{a}}_{\phi_0} \right] \right\}}{H_0} 
    \frac{ \mathcal{A} \left\{\exp \left[{\bf \hat{a}}^\dagger_{\phi_\perp} \left(I_M - H_\perp \right) {\bf \hat{a}}_{\phi_\perp}  \right] \right\}}{H_\perp} 
    \right\rangle 
    \nonumber\\
    &= \int d^2 {\bm \alpha}_{\phi_0} 
    \, \mathcal{K}({\bm \alpha}_{\phi_0},H_0) \int d^2 {\bm \alpha}_{\phi_\perp}  
    \, \mathcal{K}({\bm \alpha}_{\phi_\perp},H_\perp)  
    \, Q_{\bf r}({\bm \alpha}_{\phi_0}, {\bm \alpha}_{\phi_\perp} )
    ,
    \label{gen_func_aux2_prove}
\end{align}
which recovers Eq.~(\ref{gen_func_aux2}) of the main text with the effective Husimi function defined as:
\begin{equation}
    Q_{\text{eff}}({\bm \alpha}_{\phi_0}) =   \int d^2 {\bm \alpha}_{\phi_\perp} 
    \, \mathcal{K}({\bm \alpha}_{\phi_\perp},H_\perp)  
    \, Q_{\bf r}({\bm \alpha}_{\phi_0}, {\bm \alpha}_{\phi_\perp} )
    .
    \label{Husimi_effective_def}
\end{equation}

In this way, by substituting the squeezed state, within the low-noise approximation given in Eq.~(\ref{smss_homog}), into Eq.~(\ref{Husimi_effective_def}), we obtain the Husimi function of the effective state,
\begin{align}
    &Q_{\bf r}({\bm \alpha}_{\phi_0}, {\bm \alpha}_{\phi_\perp} )
    \equiv \frac{1}{\pi^{2M}} \big| \langle {\bm \alpha}_{\phi_0}, {\bm \alpha}_{\phi_\perp} | {\bf r} (\epsilon) \rangle \big|^2
    \nonumber\\
    & \quad =  \frac{c_{\bf r}}{\pi^{2M}} \, \exp \left( - {\bm \alpha}^\dagger_{\phi_\perp} {\bm \alpha}_{\phi_\perp} \right) \, \langle {\bm \alpha}_{\phi_0}| 
    \, \prod_{k=1}^M \exp \Big[
    \frac{S_k}{2}(\hat{a}^\dagger_{k, \phi_0})^2  + 
    S_k \sqrt{\epsilon} \, \overline{\alpha}_{k, \phi^\perp_k} \, \hat{a}^\dagger_{k, \phi_0}
    \Big] |0 \rangle \langle 0 | \prod_{k=1}^M \exp \Big[
    ~\text{h.c.}~ \Big] | {\bm \alpha}_{\phi_0} \rangle
\end{align}
and replacing in Eq.~(\ref{Husimi_effective}) we have
\begin{align}
    &Q_{\text{eff}}({\bm \alpha}_{\phi_0})
    \nonumber\\
    &=\frac{c_{\bf r}}{\pi^M} \int \frac{d^2 {\bm \alpha}_{\phi_\perp} }{\pi^M} 
    \, \frac{\exp \left( -{\bm \alpha}^\dagger_{\phi_\perp} H_\perp {\bm \alpha}_{\phi_\perp} \right)}{H_\perp}  \, \langle {\bm \alpha}_{\phi_0}| 
    \, \prod_{k=1}^M \exp \Big(
    \frac{S_k}{2}(\hat{a}^\dagger_{k, \phi_0})^2  + 
    S_k \sqrt{\epsilon} \, \overline{\alpha}_{k, \phi^\perp_k} \, \hat{a}^\dagger_{k, \phi_0}
    \Big) |0 \rangle \langle 0 | \prod_{k=1}^M \exp \Big(
    ~\text{h.c.}~ \Big) | {\bm \alpha}_{\phi_0} \rangle
    .
    \label{Husimi_effective_prove}
\end{align}
Therefore, we recover Eq.~(\ref{Husimi_effective}) by identifying $\hat{\rho}_\epsilon$ as the operator inside the integral of Eq.~(\ref{Husimi_effective_prove}) with the change of variables to the effective displacement ${\bm \beta}=\sqrt{\epsilon}\,\overline{D}_{\bf r}\,{\bm \alpha}_{\phi_\perp}$.


\newpage

\twocolumngrid
\bibliographystyle{apsrev4-2}
\bibliography{smss_dist}

\end{document}